\documentclass[aps,prb,twocolumn,amsmath,amssymb,floatfix]{revtex4-1}
\usepackage{tabularx}
\usepackage{bm}
\usepackage{euscript}
\usepackage{graphicx}
\usepackage{color}
\usepackage{amsfonts}
\usepackage{exscale}
\usepackage{amsbsy}
\usepackage{subfigure}
\usepackage{textcomp}

\newcommand{\bs}[1]{\boldsymbol{#1}}

\begin{document}

\title{Optimal T$_c$ of cuprates: role of screening and  reservoir layers}
\author{S. Raghu, R. Thomale, and T. H. Geballe}
\affiliation{Department of Physics, Stanford University, Stanford, California 94305, USA}

\date{\today}

\begin{abstract}
We explore the role of charge reservoir layers (CRLs) on the superconducting transition temperature of cuprate superconductors.  Specifically, we study the effect of CRLs with efficient short distance dielectric screening coupled capacitively to copper oxide metallic layers.  We argue that dielectric screening at short distances and at frequencies of the order of the superconducting gap, but small compared to the Fermi energy can significantly enhance T$_c$, the transition temperature of an unconventional superconductor.  We discuss the relevance of our qualitative arguments to a broader class of unconventional superconductors.  
\end{abstract}


\maketitle

\section{ introduction}
The superconducting properties of the cuprates are widely believed to be determined by the electrons in the copper-oxide (CuO$_2$) layer.  This is confirmed by experiments that have identified these electrons as the  low energy degrees of freedom\cite{Damascelli2003}.  Furthermore, the study of simplified low energy effective models, such as the single-band Hubbard model and its descendants  have provided  overwhelming evidence in favor of this view.  The robust broken symmetry phases found in the cuprates, such as antiferromagnetism and d-wave superconductivity, are unequivocally obtained as ground states of these models in appropriate limits.

However, assuming that the CuO$_2$ layers in different cuprate materials are electronically similar, the origin of substantial diversity of their optimal transition temperatures ({\it i.e.} T$_c$ at optimal doping)  is an important issue that remains poorly understood.  For instance, the optimal T$_c$ of LSCO is 40 K, whereas that of the single layer Hg-2201 compound  is more than twice as large.   It is difficult to ignore this spectacular variation, despite the fact that T$_c$ is a non-universal quantity.   Indeed, various theories have been proposed to address this issue: the prevailing view  is that alterations of the electronic structure of the CuO$_2$ layer itself must be responsible for the differences in optimal T$_c$.  For instance, the variations could occur due to differential amounts of disorder in the CuO$_2$ layer.  Another popular approach to the problem involves  relating changes in T$_c$ to differences in copper-apical oxygen bond lengths\cite{Pavarini2001, Andersen1995}, which in turn induce subtle changes in the structure of the Fermi surface\cite{Sakakibara2010, Kuroki2011,Sakakibara2012}.  

A more radical proposal\cite{Geballe1999,Geballe2006}  invokes the role of charge reservoir layers (CRLs), which are spatially separated from the CuO$_2$ layer,  in determining the optimal T$_c$.   
The CRLs are coupled to the CuO$_2$ layer capacitively.   Taken at face value, the notion that CRLs affect T$_c$ is not unreasonable: systems with CRLs such as the mercury cuprates have higher optimal T$_c$s than materials such as LSCO, which do not possess CRLs.  Moreover,  materials with different CRLs  also have substantially different optimal T$_c$s.  However, the mechanism by which the CRLs affect T$_c$ is unclear.  Here, we attempt to place the possibility that CRLs can affect T$_c$  on more firm theoretical footing. 
The work in Refs. ~  \onlinecite{Geballe1999,Geballe2006} suggested that resonant pair tunneling due to negative U centers was responsible for this enhancement.  Here, however, we take a very different approach: we argue instead that if reservoir layers were highly polarizable, they can significantly alter the effective pairing interaction (and therefore T$_c$) of unconventional superconductors.

The intuition underlying our argument can be stated as follows.  Any realistic system will always have both onsite and longer range repulsive electron interactions.  Whereas the onsite interactions (as emphasized in Hubbard-like models) reflect atomic physics at the shortest distance scales, longer range interactions reflect the solid state environment in which the low energy degrees of freedom are embedded: they are effective interactions among the essential degrees of freedom generated by ``integrating out"  the environment.  While the onsite repulsive interactions are directly responsible for the unconventional pairing, more extended repulsive interactions have the opposite effect - they weaken the scale at which pairing occurs\cite{Raghu2012}.  Therefore, if the environment (i.e. the CRL in the present context) were highly polarizable, it could act to weaken longer range interactions in the CuO$_2$ layer and therefore to enhance T$_c$.  

While the discussion here is framed largely in the context of the cuprates, we believe that the robust qualitative effects on T$_c$ emphasized here are relevant to a broader class of  materials exhibiting unconventional superconductivity.  Some of the effects described here could also be explored in artificially engineered systems consisting of hybrids of distinct parent materials.  
  
The outline of the paper is as follows.  In Section \ref{sec:pheno}, we review phenomenological arguments that lead to the effective Hamiltonian constructed in Section \ref{sec:ham}.  Section \ref{sec:sc} discusses the superconducting properties of the system of interest in various limits.  We present our conclusions and outline future directions in Section \ref{sec:disc}.

\section{Relevant phenomenology of multi-layered cuprates}
\label{sec:pheno}
\begin{figure}[t]
\begin{minipage}{0.99\linewidth}
\includegraphics[width=\linewidth]{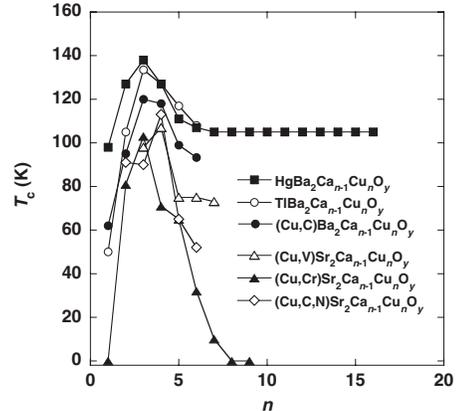}
\end{minipage}
\caption{T$_c$ in several families of multilayered cuprate superconductors  [Reproduced with permission from J. Phys. Soc. Jpn {\bf 76}, 094711 (2007)].   }
\label{tc}
\end{figure}

In this section, we discuss  phenomenological arguments that inspired us to construct and analyze the  model Hamiltonian of  Section \ref{sec:ham}.  Fig.~ \ref{tc} shows the optimal T$_c$ of several families of multi-layer cuprate superconductors, each having different CRLs.  
In these systems, each unit cell consists of n-CuO$_2$ layers stacked along the c-axis and is separated from the next by a CRL.  The CRL is separated from the outermost CuO$_2$ plane (OP) by an insulating oxide layer, which suppresses single electron tunneling between them.  Thus, the  CRL and OP form a naturally occurring ``oxide interface",  and their coupling is primarily capacitive.  
For a family of materials with the same CRL  the dependence of T$_c$ on $n$ is remarkably  universal: it increases from $n=1$, and decreases beyond an optimal  value of $n \approx 3$.  For $n > 5$, T$_c(n) \approx T_c(n=1)$.  Several theories have been proposed to address the further enhancement at $n \approx 3$ - see for instance Refs. ~\onlinecite{Kivelson2002, Chakravarty2004}.  By contrast, the point we stress here is that, as is clear from Fig.~ \ref{tc}, for families with different CRLs, the optimal T$_c$ itself  varies drastically.  
For a recent summary of the experimental data of multilayer cuprates, see Ref. ~ \onlinecite{Iyo2007}.

From simple electrostatic considerations, it follows that the OP layers are more overdoped whereas the IP layers are more underdoped.  This is also consistent with the quantum chemistry of these materials: the IP layers do not have apical oxygens and therefore are closer to half-filling than the OP layers.  Indeed, this is confirmed by the fact that antiferromagnetism has been experimentally observed in the IPs\cite{Mukuda2011}.   
With increasing $n$,  superconductivity occurs mainly on the OPs and becomes more two dimensional; the 3-layer system consisting of two OPs sandwiched by a CRL are separated from the next unit cell by a large number of intervening antiferromagnetic layers.  This point of view is strongly  substantiated by the fact that the vortex melting curves of multi-layered cuprates are two-dimensional in character, and are independent of $n$ for $n > 5$\cite{Crisan2008}.  The fact that T$_c$ at large $n$ is close to the value at $n=1$ gives us further justification for the fact that the relevant degrees of freedom as far as superconductivity is concerned,  are the CRL and OP layers.  We are led, therefore, to consider a simple model Hamiltonian built only out of these degrees of freedom.  Within the context of this simple model, we will study how the dielectric properties of the CRL can affect T$_c$ of the OP electrons.

\begin{figure}[t]
\begin{minipage}{0.99\linewidth}
\includegraphics[width=\linewidth]{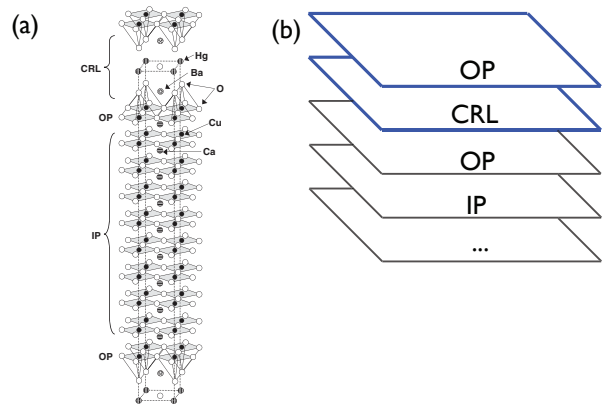}
\end{minipage}
\caption{ (a) The $n$-layer Hg cuprate consists of $n-2$ inner planes (IP), two outer planes (OP), and a CRL.   In this family, T$_c$ exhibits no variation with $n$ beyond $n>5$ and is close to T$_c(n=1)$ [Reproduced with permission from J. Phys. Soc. Jpn {\bf 76}, 094711 (2007)].  (b) For sufficiently large $n$, the inner planes are undoped and the outermost plane exhibits superconductivity.  The minimal description in this limit reduces to that of a bilayer consisting of an OP layer and a CRL as shown in (b).}
\label{multilayer}
\end{figure} 

\section{Model Hamiltonian}
\label{sec:ham}
If the CRL were to affect T$_c$ of the OP electrons by screening longer range interactions, its dielectric function must satisfy specific requirements.  Firstly, the dielectric function should be large over a range of frequencies, $k_B T_c \lesssim \omega \ll E_{\text{F}}$.  Secondly, the dielectric screening must be efficient over a broad range of momenta up to scales comparable to the inverse lattice spacing.  
A very simple way in which the CRLs could exhibit such properties is if they were metallic, or if they consisted of a liquid of dipole moments that can fluctuate and are able to screen the OP electron fluid.  
Very little is known from experiments about the dielectric properties of the CRLs and further work is needed to clarify the situation.  From a theoretical standpoint, electronic structure calculations have found the possibility that CRL bands are close to being metallic\cite{Singh1993}, and become metallic  under pressure\cite{Singh1994,Novikov1994}.  We will not speculate on the issue further here and will instead focus on the consequences that would follow {\it if} the CRLs had the dielectric properties described above.  


Let $c_{\bm k, \sigma}$ create an electron in the OP (Fig.~\ref{multilayer}b) and let $d_{\bm k, \sigma}$ create an electron in the CRL.  We express the partition function of this system as a Grassman path integral of the form
\begin{eqnarray}
Z &=& \int D \bar c_{\bm k \sigma} D c_{\bm k \sigma} D \bar d_{\bm k \sigma} D d_{\bm k \sigma} e^{- \int_0^{\beta} d \tau \mathcal L} \nonumber \\
\mathcal L &=&  \mathcal L_{\text{crl}} + \mathcal L_{\text{op}} +   \mathcal L_{\text{op-crl}}  
\end{eqnarray}
The action consists of 3 terms: 1) a contribution purely from the CRL, 2) one purely from the OP and 3) a coupling between the CRL and OP.  For purpose of illustration, we treat the CRL as a free fermion system, whereas the OP action consists of both kinetic and potential energy:
\begin{eqnarray}
\mathcal L_{\text{crl}} &=& \sum_{\bm k \sigma} \bar d_{\bm k \sigma} \left[ \partial_{\tau} + \xi_{d, \bm k} \right] d_{\bm k \sigma}  \nonumber \\
\mathcal L_{\text{op}} &=& \sum_{\bm k \sigma} \bar c_{\bm k \sigma}  \left[ \partial_{\tau} + \xi_{c, \bm k} \right] c_{\bm k \sigma} + \frac{1}{2} \sum_{\bm q} V_1(\bm q)  \hat n_{c}(\bm q) \hat n_{c }(- \bm q) \nonumber \\
\end{eqnarray}
where $\hat n_c(\bm q)$ denotes the electron density in the OP layer, and $\xi_{c, \bm k}$, $\xi_{d, \bm k}$ are the kinetic energies relative to the Fermi level of the OP and CRL layer, respectively.  Generically, the  coupling between the OP layer and the CRL  will consist of a  direct single electron tunneling matrix element $v_{\bm k}$ from the OP to the CRL as well as Coulomb interactions between them:  
\begin{eqnarray}
\mathcal L_{\text{op-crl}} &=& \sum_{\bm k} \left[ v_{\bm k} d^{\dagger}_{\bm k \sigma} c_{\bm k \sigma} + h.c.  \right] + \sum_{\bm q} V_2(\bm q) \hat n_d(- \bm q) \hat n_{c}(\bm q) 
 \nonumber \\
\end{eqnarray}
The low energy effective action for electrons in the OP layer is obtained by integrating out the electrons in the CRL.  One therefore obtains
\begin{eqnarray}
\mathcal L_{\text{eff}}  &=& \sum_{\bm k \sigma} c^{\dagger}_{\bm k \sigma} \left[ ( \partial_{\tau} + \xi_c(\bm k) ) + \Sigma(\bm k, \omega) \right] c_{\bm k \sigma} \nonumber \\ && + \sum_{\bm q} V_{\text{eff}}(\bm q) \hat n_{c}(\bm q) \hat n_{c}(- \bm q) + \cdots
\label{heff}
\end{eqnarray}
where 
\begin{eqnarray}
\Sigma(\bm k) &=&  v^*_{\bm k} (\partial_{\tau} + \xi_{d, \bm k})^{-1}  v_{\bm k} \nonumber \\
V_{\text{eff}}^{(\text{op})}(\bm q) &=& V_1(\bm q) - \frac{1}{2} V^2_2(\bm q) \chi(\bm q,\omega) 
\end{eqnarray}
and $\chi(\bm q, \omega)$ is the charge susceptibility of the CRL electrons, which in turn is related to its dielectric function.  
Here, the `` $\cdots$" involve non-linear susceptibilities that we have neglected in the spirit of RPA (linear-response): moreover, these terms produce irrelevant corrections to the effective action to be treated below.  
We note that the direct hybridization between the OP and CRL produces a self-energy correction to the Green's function of the OP electrons.  Therefore, if the CRL were sufficiently disordered, the disorder would be introduced into the OP.  The fact that an oxide barrier is present in between the OP electrons and the CRL electrons is likely to make the hybridization small, and we shall neglect it in what follows.  

In the analysis of the effective action of Eq. \ref{heff} in the next section, we will make use of the following concrete form: for the non-interacting kinetic term $H_0=\sum_{\bm k, \sigma} \xi_c(\bs{k}) c_{\bm k, \sigma}^\dagger c_{\bm k, \sigma}^{\phantom{\dagger}}$, we assume a square lattice including nearest ($t$) and next nearest neighbor hybridization ($t'$) for the OP layer, i.e. $\xi_c(\bs{k})=-2t (\cos k_x +\cos k_y) + 4 t' \cos k_x \cos k_y -\mu $.  Our goal here is to illustrate the role of longer range interactions; it will suffice to treat the interactions in Eq.~\ref{heff} as being finite-ranged as in an extended-Hubbard model:
\begin{eqnarray}
&&H=H_0+H_{\text{int}}, \nonumber \\
&&H_{\text{int}}= U \sum_{i,\sigma} c_{i,\sigma}^\dagger c_{i,\sigma} + \frac{U_1}{2}\sum_{\langle i,j \rangle, \sigma, \sigma'} n_{i,\sigma}n_{j.\sigma'} \nonumber \\
&&\phantom{ddddd.}+\frac{U_2}{2}\sum_{\langle \langle i,j \rangle \rangle, \sigma, \sigma'} n_{i,\sigma}n_{j.\sigma'},\label{htotal}
\end{eqnarray}
where $n_{i,\sigma}$ denotes the OP fermionic occupation operator of spin $\sigma$ at site $i$ of the square lattice. Within this description, the effect of capacitive screening from the CRL is to change the ratio $U_1/U$ and $U_2/U$ in~\eqref{htotal}. This will be the starting point for our subsequent analysis.

\section{Superconductivity} 
\label{sec:sc}
In this section, we study the pairing scale associated with the model~\eqref{htotal} in several different limits and find the same qualitative trend in each case.  Firstly, in the limit where the interactions within the OP layers are weak compared to the kinetic energy, we  discuss asymptotically exact  perturbative renormalization group results\cite{Raghu2010,Raghu2012}.   In this limit, the normal state is a well-behaved Fermi liquid and superconductivity is the only instability of this system.  We next consider the intermediate coupling regime. In this regime, there are no small parameters, and therefore are no well-controlled methods with which to attack the problem.  However, there is an unbiased approach known as the functional renormalization group (fRG), which treats all possible ordering tendencies on equal footing.    
This approach has provided us with important guidance and intuition regarding the pairing tendencies of the cuprates and pnictide superconductors~\cite{polchinski84npb269,shankar94rmp129,zanchi-00prb13609,halboth-00prb7364,honerkamp-01prb035109,Thomale2009,Wang2009,Thomale2011}.
 Lastly, we consider the limit of strong coupling in which the low energy behavior of the system is dictated by the proximity to a Mott insulating phase.  In this limit, there do not exist any controllable analytic treatments.  However, there are approximate treatments based on the idea of spin-charge separation and electron fractionalization.  These methods, known as slave particle mean-field theories are the simplest means of obtaining a system with a large Fermi surface, but with a low superfluid density and provides us with physical intuition as to the nature of the underdoped cuprates\cite{Ruckenstein1987,Baskaran1988, Kotliar1988,Ubbens1994,Lee1997,Lee1998, Lee2006}.  We stress below that in all three limits, when longer range interactions are screened, the scale at which superconductivity develops is enhanced.  

\subsection{Asymptotically exact perturbative renormalization group treatment at weak coupling}
\label{weak}
In the weak-coupling limit $U \ll W$, where $W$ is the kinetic bandwidth, superconductivity develops out of a well-behaved Fermi liquid, and is the only generic instability of the system, as was first demonstrated in Ref. ~ \onlinecite{Kohn1965}.    
In a modern formulation of this problem\cite{Raghu2010}, the problem is solved in two stages.  In the first stage, one first introduces an unphysical, artificial cutoff $\Omega_0$ and integrates out all modes with energy $E > \Omega_0$.  The cutoff  is chosen so that $W \exp{(-1/\rho U)} \ll \Omega_0 \ll U^2/W$ and the modes can be integrated out perturbatively in $U/W$ to as high a degree of precision that is desired.  When this is accomplished, the remaining electronic modes lie parametrically close to the Fermi surface, and induced attractive interactions due to the particle-hole continuum   are generated in the low energy effective action.  In the second stage, the perturbative RG flows of the resulting effective action are computed and marginally relevant couplings destabilize the Fermi liquid to produce the superconducting instability below a scale
\begin{equation}
T_c \sim W e^{-1/  a U^2},
\end{equation}
where $a$ is a number of order unity that involves the entire spectrum of particle-hole fluctuations.  
These results are asymptotically exact: the pairing scale is expressed as an asymptotic series in the bare couplings and the leading corrections to the expression above can easily be computed.  

When  only onsite interactions are present, the $s$-wave state in which Cooper pairing occurs on the same lattice site is disfavored; the induced attractive interactions due to the particle-hole fluctuations, however, favor unconventional superconducting states.   For electrons in a nearly half-filled tetragonal lattice, the predominant instability is towards  $d_{x^2-y^2}$ superconductivity\cite{Raghu2010}. 

When longer-range interactions are present, the induced attraction generated from the particle-hole continuum must overcome the bare repulsive forces in order for unconventional superconductivity to occur.  The interplay between these bare long range repulsive forces and the induced attractive interactions results in a rich phase diagram that was studied in detail in Ref. ~\onlinecite{Raghu2012}.  Here, we summarize the main conclusions of this work.  Since we are working in the limit as $U, U_1, U_2, \cdots$ 
all approach zero, and since the effective interactions form an asymptotic, not a convergent, series in these interactions, the result sensitively depends on the way in which the limit is taken.  If we set $U_1$ to approach zero as $U_1 \sim U^2$, there is direct competition between the induced attractive interactions $\mathcal O(U^2)$ and $U_1$, which exponentially weakens the pairing strength associated with the d$_{x^2-y^2}$ superconductor.  However, the superconducting state itself remains robust for a finite range of $U_1$.  On the other hand, if $U_1$ scales as $U_1 \sim U$ in this limit, induced attractive interactions cannot overcome the bare repulsion and the result is that the d$_{x^2-y^2}$ pairing  is completely lost.  
In general, as the range of the repulsive interaction is increased, the separation between Cooper pairs also increases\cite{kieselarxiv}.

In summary, in the limit $U \ll t$, longer range repulsive interactions substantially weaken the pairing scale at which d$_{x^2-y^2}$ superconductivity occurs; therefore, if a polarizable medium were to be placed in proximity to the metal, the superconducting pairing strength would be exponentially enhanced.

\subsection{Functional renormalization group treatment at intermediate coupling}
\label{inter}
\begin{figure}[b]
\begin{minipage}{0.90\linewidth}
\includegraphics[width=\linewidth]{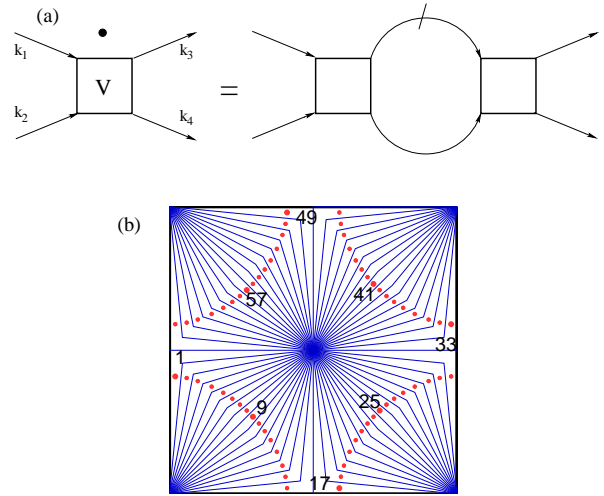}
\end{minipage}
\caption{(color online) (a) FRG parquet flow equation of the two-particle vertex $V_{\Lambda}$. The dotted derivative denotes $\frac{\partial}{\partial \Lambda} V^{\Lambda}$, the dashed internal line is the single scale propagator $\frac{\partial}{\partial \Lambda} G_{0, \Lambda}$. (b) Patching scheme ($N=64$, counterclockwise) for a typical cuprate Fermi surface at $t'=0.3t$ and filling $n=0.98$.}
\label{picfrg1}
\end{figure}
In the weak-coupling limit treatment above, there generically isn't any competition between unconventional pairing and non-superconducting orders, unless one fine-tunes the band structure.  In the intermediate coupling regime, longer range interactions have two effects: they lower the pairing scale, as in the weak-coupling case, but they also enhance competing orders, as we discuss below.  In this subsection, we discuss both effects  and show that both act to lower the transition temperature.  
   
As an adaptation of the Wilsonian RG for interacting Fermi systems~\cite{polchinski84npb269,wegner94,shankar94rmp129}, the functional renormalization group (FRG) of the two-dimensional Hubbard model has been introduced by different groups~\cite{zanchi-00prb13609,halboth-00prb7364,honerkamp-01prb035109,metzner-12rmp299}. 
Within FRG, the flow of the two-particle vertex function $V^{\Lambda}(\bs{k}_1,\bs{k}_2, \bs{k}_3, \bs{k}_4)$ is studied as a function of the energy cutoff parameter $\Lambda$, where $\bs{k}_1$ and $\bs{k}_2 $ ($\bs{k}_3$ and $\bs{k}_4$) denote the ingoing (outgoing) particles (Fig.~\ref{picfrg1}a). $\Lambda$ denotes the energy scale up to which high energy modes have been integrated out to provide an effective two-particle interaction vertex. Its initial conditions are given by the bare interaction in~\eqref{htotal}. The largest scale $\Lambda_c$ at which the RG flows break down will signal a Fermi surface instability in a given interaction channel.  While the absolute magnitude of $\Lambda_c$ can differ from T$_c$,  the variation of $\Lambda_c$ upon changing system parameters is indicative of relative variations of $T_c$.  
The momenta $\bs{k}_i$ entering the vertex function $V^{\Lambda}$ are constrained to take $N$ finite values in the Brillouin zone, which is divided into patches (Fig.~\ref{picfrg1}b). The flow equation of the two-particle vertex then becomes a system of coupled $N^3$ integro-differential equations, which are then solved numerically (Fig.~\ref{picfrg1}a)~\cite{metzner-12rmp299}. 
As we ignore all self-energy effect in the flow, the single-particle Green function $G_{0,\Lambda}=\frac{C(\Lambda)}{i\omega - \epsilon(\bs{k})}$ only changes with respect to the cutoff function $C(\Lambda)$~\cite{metzner-12rmp299} .  
%

The main approximations of the FRG are given by (i) neglecting the self energy corrections imposed by the two-particle vertex, (ii) discarding the frequency dependence of the vertex, (iii) limited momentum resolution due to the finite patch number $N$, and (iv) neglecting higher order diagrams. 
It is instructive to see how the weak coupling ansatz in Section~\ref{weak} and the FRG are connected: in the limit of infinitesimal $U$, the approximations (i), (ii), and (iv) become irrelevant, and the precision along (iii) can be enhanced according to a regular numerical momentum integration in the Brillouin zone.

For intermediate coupling, superconductivity generically competes with density-wave type instabilities: 
at high $\Lambda$, the main feature emerging in the renormalization group flow of the interaction vertex are particle-hole fluctuations, which then seed superconducting fluctuations~\cite{Kohn1965}.  The phase diagram for local Hubbard interactions has been obtained in Ref.~\onlinecite{honerkamp-01prb035109}. Nearby nesting, SDW order is dominant for sufficiently large $U$, while $d_{x^2-y^2}$ superconductivity wins either by doping or enhancement of $t'$.
\begin{figure}[t]
\begin{minipage}{0.90\linewidth}
\includegraphics[width=\linewidth]{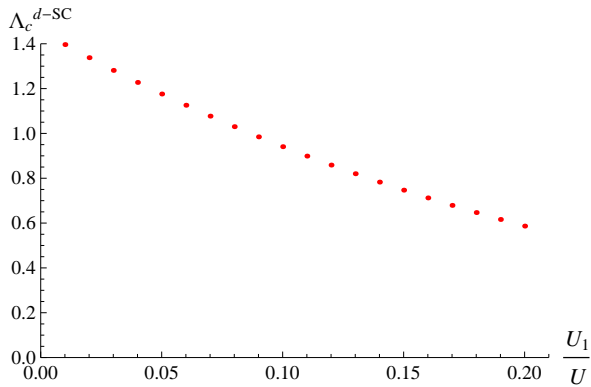}
\end{minipage}
\caption{(color online) Critical scale $\Lambda_c^{\text{SC}}$ as a function of $U_1/U$, with fixed parameters $U_=2t$, $n=0.98$, $t'=0.3t$. $\Lambda_c$ decreases significantly for finite $U_1$.}
\label{picfrg2}
\end{figure}

$d_{x^2-y^2}$-wave superconductivity stays the leading superconducting instability for intermediate coupling when we consider $U$ and $U_1$ interactions according to~\eqref{htotal}. 
$\Lambda_c$ decreases as $U_1$ is enhanced (Fig.~\ref{picfrg2}) because the long-range part of Coulomb interactions considerably affects the critical scale of unconventional superconductivity at intermediate coupling. The sensitivity of $\Lambda_c$ to longer range interactions is strong when competing orders are present in the particle hole channel: in the vicinity of half filling, enhancing $U$ can provide a change from SC to SDW order; furthermore, $U_1/U>0.6$ drives the system into a charge density wave (CDW) phase whose $\Lambda_c$ increases for higher $U_1$. In the overdoped regime, competing orders in the particle hole channel are generally suppressed due to absence of Fermi surface nesting. However, we still find $\Lambda_c^{\text{SC}}$ to slightly decrease upon enhancing $U_1$ which is due to reminiscent competing particle-hole fluctuation effects.

%

\subsection{Strong coupling}
We now turn to the limit where the onsite and extended electron-electron interactions are much larger than their kinetic energy.  Specifically, we start by investigating~\eqref{htotal}, and assume $t'=0$ for simplicity.
In the strong coupling limit, we take the onsite interaction $U/t \rightarrow \infty$.  In this limit, close to half-filling, one projects the system onto the lower Hubbard band consisting only of singly occupied sites.  The mixing between the Hubbard bands are eliminated via a canonical transformation and a $t/U$ expansion, which leads to the following effective Hamiltonian\cite{Auerbach, Fazekas}:
\begin{eqnarray}
H_{eff} &=& \mathcal P \left[ - t\sum_{\langle i j \rangle} c^{\dagger}_{i \sigma} c_{j \sigma} + \sum_{i j } V_{ij}(n_i - 1)(n_j - 1) \right] \mathcal P \nonumber \\
&+&  J \sum_{\langle i j \rangle } \left[ \bm S_i \cdot \bm S_j - \frac{1}{4} n_i n_j \right], 
\end{eqnarray}
where $V_{ij}$ takes on the values $U_1$ for nearest neighbor and $U_2$ for next nearest neighbor repulsion and $\mathcal P$ is the Gutzwiller projection operator onto the subspace without any doubly occupied sites.  The exchange coupling $J = 4t^2/(U-U_1)$ depends weakly on the longer range interactions and sets the scale for the low energy spin dynamics of the system near half-filling.    It is important to stress that unlike the onsite interactions, the extended interactions {\it do} contribute to the dynamics of the lower Hubbard band.  This will certainly hold true in the approximate, yet physically motivated treatments of this problem that we consider below.  

A method that is frequently used to handle the projection involves the introduction of auxiliary `slave' particles $f_{i \sigma}, b_i$ defined via
\begin{equation}
c^{\dagger}_{i \sigma} = b_i f^{\dagger}_{i \sigma}.
\end{equation}
The physical electron operator is written as a product of a charged bosonic operator $b_i$ and a neutral fermion operator $f_{i \sigma}$.  The projection is implemented via the constraint 
\begin{equation}
\sum_{\sigma} f^{\dagger}_{i \sigma} f_{i \sigma} + b^{\dagger}_i b_i = 1
\end{equation}
so that, {\it e.g.} matrix elements of operators involving the product $n_{f i} n_{b,i}$, where $n_{fi}(n_{bi})$ are the slave fermion(boson) densities, vanish in the projected Hilbert space.  The constraint is implemented by introducing a $U(1)$ gauge field whose fluctuations are neglected in the simplest mean-field theories.    

The mean-field treatment involves a series of approximations.   Firstly,  the constraint is satisfied only on average, i.e. $  \sum_{\sigma}  \langle f^{\dagger}_{i \sigma} f_{i \sigma} \rangle+ \langle b^{\dagger}_i b_i \rangle = 1
$.  Secondly, in the superconducting phase, the slave bosons  condense and exhibit off-diagonal long-range order.  Therefore, the operator $b_i$ is replaced by a c-number $ \langle b_i \rangle \rightarrow \sqrt{\delta}$, where $\delta$ is the concentration of holes, and near half-filling, $\delta \ll 1$.  The resulting Hamiltonian does not involve the projection operator $\mathcal P$ and  can be written in the form
\begin{eqnarray}
H =&& -t \delta \sum_{\langle i j \rangle \sigma} \left( f^{\dagger}_{i \sigma} f_{j \sigma} + h.c. \right) + J \sum_{\langle i j \rangle } \left[ \bm S_i \cdot \bm S_j - \frac{1}{4} n_i n_j \right] \nonumber \\
&& +  \sum_{i j } V_{ij}(n_i-1) (n_j-1)
\end{eqnarray}
Here, density and spin operators involve only the spinon degree of freedom and the subscript $``f"$ on the density operator has been eliminated for simplicity.  As a consequence, the longer range interactions are ``felt" also by spinons, despite the fact that they represent neutral particles.  

The spinon Hamiltonian above still involves quartic terms: the last approximation involves treating it variational via an unrestricted Hartree-Fock approximation.  Specifically, we introduce a trial Hamiltonian of the form 
\begin{equation} 
H_{\text{tr}}  = \sum_{\bm k \sigma} \epsilon_{\bm k} f^{\dagger}_{\bm k \sigma} f_{\bm k \sigma} + \frac{1}{2} \sum_{\bm k \sigma \sigma'} \left[ \tilde \Delta_{\bm k}f^{\dagger}_{\bm k \sigma} (i \tau^y_{\sigma \sigma'} )f^{\dagger}_{- \bm k \sigma'} + h.c. \right]
\end{equation}
where $\tilde \Delta_{\bm k}$ is a variational parameter corresponding to the ``gap" due to pre-formed spinon pairs.  The variational free energy $F_0 = F_{\text{tr}} + \langle H - H_{\text{tr}} \rangle_{\text{tr}}$ is extremized, which leads to the following self-consistent gap equation for $\tilde \Delta_{\bm k}$:
\begin{equation}
\frac{1}{J-U_1} = \sum_{\bm k} \left( \cos{k_x} - \cos{k_y} \right)^2 \frac{1}{2 E_{\bm k}} \tanh{\left[ \frac{\beta E_{\bm k}}{2} \right] }
\end{equation}
where $E_{\bm k} = \sqrt{\epsilon_{\bm k}^2 + \tilde \Delta_{\bm k}^2}$.  
The physical order parameter corresponding to the superconducting state is $\Delta_{ij} = \delta \tilde \Delta_{ij} $.  From the gap equation, it follows that the magnitude of the spinon gap decreases linearly as the nearest neighbor repulsion $U_1$ increases.  However, since the gap has  d$_{x^2-y^2}$ symmetry, it  is relatively unaffected by the second-neighbor repulsion.  Similar results were found in DMRG studies of extended Hubbard models on ladder systems\cite{Raghu2012}.  An additional effect of longer range interactions is that they disfavor superconductivity by enhancing the tendency towards competing orders such as flux phases and density-wave orders: we have observed this competition in an explicit Hartree-Fock analysis of the spinon Hamiltonian above and will discuss our results elsewhere.  

The longer range interactions have another significant consequence, which is purely quantum mechanical in origin.  They suppress charge fluctuations and in turn enhance phase fluctuations\cite{Doniach1990}.    This is most easily seen from an alternative mean-field decoupling of the extended $t$--$J$ Hamiltonian above in which the system is expressed in terms of the slave boson degrees of freedom.  When longer range interactions are repulsive (attractive), the superfluid density decreases(increases) linearly with $U_1, U_2 \cdots$.  In this regime, longer range components of the electron interactions affect T$_c$ linearly.  Thus even in the strong coupling limit which is relevant to the underdoped cuprates, 
screening from a proximate polarizable media can act to raise T$_c$.

\section{Discussion and Conclusion}
\label{sec:disc}
There are many factors that play a role in optimizing T$_c$: after all, since it is a non-universal quantity, it will depend  sensitively on small variations of the microscopic properties of a material.  To make quantitative predictions of T$_c$, a complete understanding of the microscopic pairing mechanism is required, taking into account all material-specific details.  Clearly, this is an impossible task at present.  Alternatively, by searching for robust qualitative phenomena that depend on microscopic physics in a simple parametric fashion, and by looking at relative trends of pairing strengths, one could in principle uncover new strategies for optimizing T$_c$ in existing unconventional superconductors.  

Indeed, this type of pursuit has led us to understand how altering the properties of the   CuO$_2$ layers can optimize T$_c$.  Examples include ideas that focus on the relative influence of the pairing scale and superfluid density\cite{Emery1995,Kivelson2002,Berg2008,Goren2009} in optimizing T$_c$, as well as on the effect of delocalizing Cooper pairs in layered structures via interlayer tunneling\cite{Chakravarty2004} .  These scenarios may have much to do with the universal features of the T$_c$ curves in Fig.~ \ref{tc}.  However, these ideas focus on the properties of the CuO$_2$ layers and  cannot address the question of why, for instance the optimal T$_c$ of Hg-based cuprates are much larger than LSCO and LBCO which do not have CRLs.  

By similarly focusing on robust qualitative effects on T$_c$, we have shown that highly polarizable media, when coupled capacitively to a metal, can act to raise the transition temperature of an unconventional superconductor.  We have shown this by considering the problem in various limits and arguing within the phenomenological context of extended lattice Hubbard models, that the longer range components of the repulsive interactions are always detrimental to d-wave superconductivity.  
We stress here that this result does not require us to invoke a sharply defined bosonic pairing `glue'.  

At present, very little is known experimentally about the CRLs.  It would therefore be of much interest in this context to obtain the dielectric properties of the CRL using resonant x-ray spectroscopic methods.  The dielectric properties of CRLs can also be investigated theoretically in first principles calculations, which we believe are worth undertaking.  If the CRLs did possess the required dielectric properties, the behavior of T$_c$ would be non-monotonic as a function of uniaxial pressure, applied along the c-axis.  As the CRLs move closer to the OP layers, T$_c$ would increase due to the mechanism that we discussed above.  However, if the CRLs were too close to the OP layers, even onsite interactions would be screened; this in turn would lower T$_c$.  


While we have been motivated here by material specific considerations involving the role of the charge reservoir layers, our findings are likely to apply to a broader class of unconventional superconductors.  With this in mind, it will be vital to explore these ideas in artificially engineered interfaces between metals and polarizable media where the complications due to material properties are less pronounced.  An example is an interface between amorphous dipolar liquids (i.e. amorphous mixtures of ferroelectric and anti-ferroelectric subsystems) and correlated metals.  A  particularly exciting example is the case of a 2DEG with metallic gates where screening due to the gates may enable the observation of unconventional superconductivity.    In this context, by tuning the ratio of the bandwidths of the 2DEG relative to that of the metallic gates, it will be possible, in principle, to {\it overscreen} the  Coulomb interactions resulting in attraction.  We shall pursue these studies in a future publication.  

By contrast, the effects of screening on T$_c$ that we consider here are unlikely to apply to conventional electron-phonon superconductors.  Even in these systems, the attractive interaction due to electron-phonon coupling, $\lambda$,  must overcome the renormalized Coulomb pseudopotential $\mu^*$.  However,  the range of the Coulomb interaction has virtually no consequence for T$_c$: the retarded interaction is usually local in real space and is insensitive to any subtle momentum-dependence of $\mu^*$.  Furthermore, the quantity $\mu^*$ depends mainly on the ratio of $\omega_D/E_F$ and hardly depends on the bare Coulomb interaction.  It therefore follows that the screening mechanisms we consider here would likely not affect its magnitude.

{\bf Acknowledgements}
We thank A.~V.~Chubukov, D.~Orgad and T.~P.~Devereaux for discussions.  We are especially grateful to S.~Chakravarty and S.~Kivelson for inspiring discussions.   RT thanks M.~Kiesel and C.~Platt for collaborations on related works. SR was supported by the Alfred P. Sloan Foundation and by the LDRD program at SLAC.  RT is supported by an SITP fellowship by Stanford University.  THG is supported in part by the  DOE Office of Basic Energy Sciences, Materials Sciences and Engineering Division, under Contract DE-AC02-76SF00515.We thank the Aspen Center for Physics where part of this work was carried out.  

\bibliography{screening}

\end{document}